\documentclass[
letterpaper,
reprint,           
twocolumn,
superscriptaddress,
amsmath,           
amssymb,           
aps,               
prl,               
notitlepage,       
longbibliography,  
floatfix,          
nofootinbib,
]{revtex4-1}

\usepackage{subfigure}
\usepackage{amsmath}
\usepackage{lipsum}
\allowdisplaybreaks[4]
\usepackage{cancel}
\usepackage{extarrows}
\usepackage{tensor}     
\usepackage{float}
\usepackage[final]{graphicx}   
\usepackage[
colorlinks=true,        
citecolor=blue,         
linkcolor=blue,         
urlcolor=blue           
]{hyperref}             
\usepackage{bm}         
\usepackage{xcolor}     
\usepackage{lipsum}
\usepackage{color}      
\usepackage[utf8]{inputenc} 
\usepackage[section]{placeins} 
\usepackage{appendix}
\usepackage{units}
\usepackage[capitalise]{cleveref}
\usepackage{units}

\newcommand{\nc}{\newcommand*}

\nc{\xbar}{\bar{x}}
\nc{\rhoeq}{\rho_{\mathrm{eq}}}
\nc{\zeq}{z_{\mathrm{eq}}}
\nc{\tla}{\tilde{\lambda}}
\nc{\bt}{\beta}
\nc{\dt}{\delta}
\nc{\Dt}{\Delta}
\nc{\vj}{\vec{j}}
\nc{\vl}{\vec{l}}
\nc{\hx}{\hat{x}}
\nc{\hy}{\hat{y}}
\nc{\bj}{\bm{j}}
\nc{\mJ}{\mathcal{J}}
\nc{\mP}{\mathcal{P}}
\nc{\Msun}{M_\odot}
\nc{\app}{\approx}
\nc{\av}[1]{\langle #1 \rangle}
\nc{\eq}[1]{Eq.~\eqref{#1}}
\nc{\al}{\alpha}
\nc{\Xstar}{X_{\ast}}
\nc{\fpbh}{f_{\mathrm{pbh}}}
\nc{\vth}{\vec{\theta}}
\nc{\vla}{\vec{\lambda}}
\nc{\vd}{\vec{d}}
\nc{\Mmin}{M_{\mathrm{min}}}
\nc{\rmd}{\mathrm{d}}
\nc{\mmin}{{m_{\mathrm{min}}}}
\nc{\mmax}{{m_{\mathrm{max}}}}
\nc{\mR}{\mathcal{R}}
\nc{\tmR}{\tilde{\mathcal{R}}}
\nc{\s}{\sigma}
\nc{\ogw}{\Omega_{\mathrm{GW}}}
\nc{\addref}{[\textcolor{red}{add ref}] }
\nc{\Om}{\Omega}
\nc{\gm}{\gamma}
\nc{\Gm}{\Gamma}
\nc{\gpcyr}{\mathrm{Gpc}^{-3}\,\mathrm{yr}^{-1}}
\nc{\Eq}[1]{Eq.~\eqref{#1}}
\nc{\Fig}[1]{Fig.~\ref{#1}}
\nc{\Table}[1]{Table~\ref{#1}}
\nc{\lvc}{LIGO/Virgo} 
\nc{\Sec}[1]{Sec.~\ref{#1}}
\nc{\eg}{\textit{e.g.~}}
\nc{\SNR}{\mathrm{SNR}}
\nc{\be}{\mathbf{\epsilon}}
\nc{\bn}{\mathbf{n}}
\nc{\bd}{\mathbf{d}}
\nc{\ba}{\mathbf{a}}
\nc{\eps}{\epsilon}
\nc{\bnu}{\mathbf{\nu}}
\nc{\mb}{\mathbf}
\nc{\bbt}{\mathbf{t}}
\nc{\bth}{\mathbf{\theta}}
\nc{\bep}{\mathbf{\epsilon}}
\nc{\uni}{\mathrm{U}}
\nc{\logu}{\operatorname{\mathrm{log-U}}}
\nc{\RN}{\mathrm{RN}}
\nc{\BN}{\mathrm{BN}}
\nc{\GN}{\mathrm{GN}}
\nc{\mcN}{\mathcal{N}}
\nc{\GWB}{\mathrm{GW}}
\nc{\yr}{\mathrm{yr}}
\nc{\Am}{\mathcal{A}}
\nc{\Dm}{\mathcal{D}}
\nc{\Hm}{\mathcal{H}}
\nc{\sovast}{Soviet Ast.}

\nc{\mrm}{\mathrm}
\nc{\BE}{B\scriptsize{AYES}\normalsize{E}\scriptsize{PHEM}\normalsize  }

\nc{\Ostgw}{\Omega_{\mathrm{GW}}^{\mathrm{ST}}}
\nc{\Ottgw}{\Omega_{\mathrm{GW}}^{\mathrm{TT}}}
\nc{\Ovlgw}{\Omega_{\mathrm{GW}}^{\mathrm{VL}}}
\nc{\Oslgw}{\Omega_{\mathrm{GW}}^{\mathrm{SL}}}
\nc{\cosxi}{\beta}

\nc{\gmPL}{\gamma_{\mathrm{PL}}}
\nc{\APL}{A_{\mathrm{PL}}}

\def\({\left(}
\def\){\right)}
\def\[{\left[}
\def\]{\right]}

\def\e{\begin{equation}}
\def\q{\end{equation}}
\def\m{\begin{eqnarray}}
\def\n{\end{eqnarray}}
\nc{\red}[1]{\textcolor{red}{#1}}

\begin{document}

\title{Constraints on the velocity of gravitational waves from NANOGrav 15-year data set}

\author{Yan-Chen Bi}
\email{biyanchen@itp.ac.cn}
\affiliation{CAS Key Laboratory of Theoretical Physics, 
    Institute of Theoretical Physics, Chinese Academy of Sciences,Beijing 100190, China}
\affiliation{School of Physical Sciences, 
    University of Chinese Academy of Sciences, 
    No. 19A Yuquan Road, Beijing 100049, China}
\author{Yu-Mei Wu}
\email{ymwu@ucas.ac.cn}
\affiliation{School of Physical Sciences, 
    University of Chinese Academy of Sciences, 
    No. 19A Yuquan Road, Beijing 100049, China}
\affiliation{School of Fundamental Physics and Mathematical Sciences, Hangzhou Institute for Advanced Study, UCAS, Hangzhou 310024, China}
\author{Zu-Cheng~Chen}
\email{zucheng.chen@bnu.edu.cn}
\affiliation{Department of Astronomy, Beijing Normal University, Beijing 100875, China}
\affiliation{Advanced Institute of Natural Sciences, Beijing Normal University, Zhuhai 519087, China}
\affiliation{Department of Physics and Synergistic Innovation Center for Quantum Effects and Applications, Hunan Normal University, Changsha, Hunan 410081, China}
\author{Qing-Guo Huang}
\email{huangqg@itp.ac.cn}
\affiliation{CAS Key Laboratory of Theoretical Physics, 
    Institute of Theoretical Physics, Chinese Academy of Sciences,Beijing 100190, China}
\affiliation{School of Physical Sciences, 
    University of Chinese Academy of Sciences, 
    No. 19A Yuquan Road, Beijing 100049, China}
\affiliation{School of Fundamental Physics and Mathematical Sciences, Hangzhou Institute for Advanced Study, UCAS, Hangzhou 310024, China}


\begin{abstract}
General relativity predicts that gravitational waves propagate at the speed of light. 
Although ground-based gravitational-wave detectors have successfully constrained the velocity of gravitational waves in the high-frequency range, extending this constraint to the lower frequency range remains a challenge. In this work, we utilize the deviations in the overlap reduction function for a gravitational-wave background within pulsar timing arrays to investigate the velocity of gravitational waves in the nanohertz frequency band. By analyzing the NANOGrav 15-year data set, we obtain a well-constrained lower bound for the velocity of gravitational waves that $v \gtrsim 0.87\,c$, where $c$ is the speed of light.

\end{abstract}
\maketitle

\section{Introduction} 
General relativity (GR) predicts three significant characteristics of gravitational waves (GW): propagating at the speed of light, two tensor polarization modes, and quadrupole radiation. While extensive research has been conducted on the latter two characteristics~\citep{Wu:2021kmd,Chen:2021wdo,Chen:2021ncc,Bernardo:2023zna,NANOGrav:2021ini,NANOGrav:2023gor}, studies often tend to focus on scenarios involving a non-zero graviton mass when it comes to propagation~\citep{Wu:2023pbt,Wu:2023rib}, thereby overlooking a generic modification of the velocity of GWs itself. 

Ground-based detectors, such as LIGO, Virgo and KAGRA, have been observing deterministic GW signals at high-frequency (Hz $\sim$ kHz) from the final merger of compact binary systems~\citep{LIGOScientific:2016aoc}.
These observations have significantly advanced our understanding of gravity~\citep{LIGOScientific:2017v1,LIGOScientific:2017v2,Isi:2017fbj,LIGOScientific:2017bnn}.
Notably, the event GW170817 has constrained the propagation velocity of GWs as $\vert 1-v \vert \lesssim 10^{-15}$ at the frequency of $f \sim 100{\rm Hz}$~\citep{LIGOScientific:2017v1,LIGOScientific:2017v2}.
However, the velocity constraint at high frequencies may not necessarily apply to the lower frequency range. Therefore, it is essential to scrutinize the constraints on velocity from a lower frequency band, which are accessible by pulsar timing arrays (PTAs).


PTAs are optimal for detecting the stochastic gravitational-wave background (SGWB) at nHz by monitoring the times of arrival (TOAs) of radio pulses emitted by a set of millisecond pulsars over decades. Recently, the North American Nanohertz Observatory for Gravitational Waves (NANOGrav)~\citep{NANOGrav:2023hde,NANOGrav:2023gor}, the European PTA (EPTA) align with the Indian PTA (InPTA)~\citep{EPTA:2023sfo,EPTA:2023fyk}, the Parkes PTA (PPTA)~\citep{Zic:2023gta, Reardon:2023gzh}, and the Chinese PTA (CPTA)~\citep{Xu:2023wog} have announced evidence for a stochastic signal consistent with the Hellings-Downs correlations~\citep{Hellings:1983fr}, pointing to the SGWB origin of this signal. 

The SGWB serves a valuable tool for revealing variations in the phase velocity of GWs.
These variations, predicted by several modified gravity theories~\citep{Schumacher:2023jxq,CarrilloGonzalez:2022fwg,Ezquiaga:2021ler,deRham:2019ctd}, can impact the overlap reduction function (ORF) in PTAs~\citep{Liang:2023ary}, providing an effective diagnostic for deviations from GR. Previous attempts to constrain the velocity using the SGWB \citep{Bernardo:2023mxc,Bernardo:2023zna} have been flawed as they only fit the spatial correlations while disregarding the information provided by the GW energy density. In this work, we conduct a comprehensive investigation by considering both the spatial correlations and energy density spectrum of the SGWB.

In this paper, we utilize the NANOGrav 15-year data set to impose constraints on the velocity of the GW via the investigation of the SGWB. 
It is worth noting that we do not delve into the distinction between phase velocity and group velocity \citep{Liang:2023ary, Bernardo:2022rif}. Our analysis uncovers a novel constraint on the GW velocity. This constraint is robust for lower values but appears weaker at higher values. To be more precise, the posterior sharply truncates when the velocity is subluminal, while it remains relatively flat when the velocity is superluminal. This outcome suggests that the available data can only discern a lower limit for the velocity of GW. Throughout this paper, we employ geometric units with $c = G = \hbar = 1$.
The rest of the paper is organized as follows. In Sec.~\ref{sec:orf}, we review the ORF as a function of GW velocity for an SGWB. In Sec.~\ref{data}, we describe the data and methodology for the analyses. Finally, in Sec.~\ref{result}, we present the results and discuss their implications.

\section{\label{sec:orf}Overlap Reduction Function} 
We now briefly review the calculation of the ORF when GWs propagate at a constant speed $v$.
We adopt a parameterized dispersion relation as
\e
\omega = v k ,
\q
where $\omega$ is the angular frequency, and $k$ is the wave number. It's worth noting that, in this expression, both phase velocity and group velocity are identical and equal to 
$v$, thus avoiding any confusion between the two. After introducing this relationship, the mode function of the GW plane wave is given by 
\e
h_{ij}\(t - \frac{1}{v} \hat{k} \cdot \vec{x}\) = \int df h_{ij}\(f, \frac{1}{v} \hat{k}\) e^{i 2\pi f \(t - \frac{1}{v} \hat{k} \cdot \vec{x}\)} ,
\q
where the velocity $v$ 
encodes the deviation from GR. When setting $v = 1$, it reduces to the GR case.

An SGWB causes delays in each pulsar’s TOAs (or in other word timing residuals) in a characteristic spatial correlated way. The corresponding timing-residual cross power spectral density between any two pulsars, $a$ and $b$, can be modeled by a power-law form 
\e
S_{ab}(f) = \Gamma_{ab} \frac{A^2_{\rm GWB}}{12 \pi^2}\(\frac{f}{f_{\rm yr}} \)^{-\gamma} f^{-3}_{\rm yr} ,
\q
where $A_{\rm GWB}$ is the amplitude of the SGWB at the reference frequency $f_{\rm yr} = 1/{\rm year}$, $\gamma$ is the spectral index of SGWB, and $\Gamma_{ab}$ is the ORF that describes average correlations between pulsars $a$ and $b$ in the array as a function of the angular separation between them. 
Note that only the tensor mode is considered throughout this work.


\begin{figure}[tbp]
	\centering
	\includegraphics[width=0.5\textwidth]{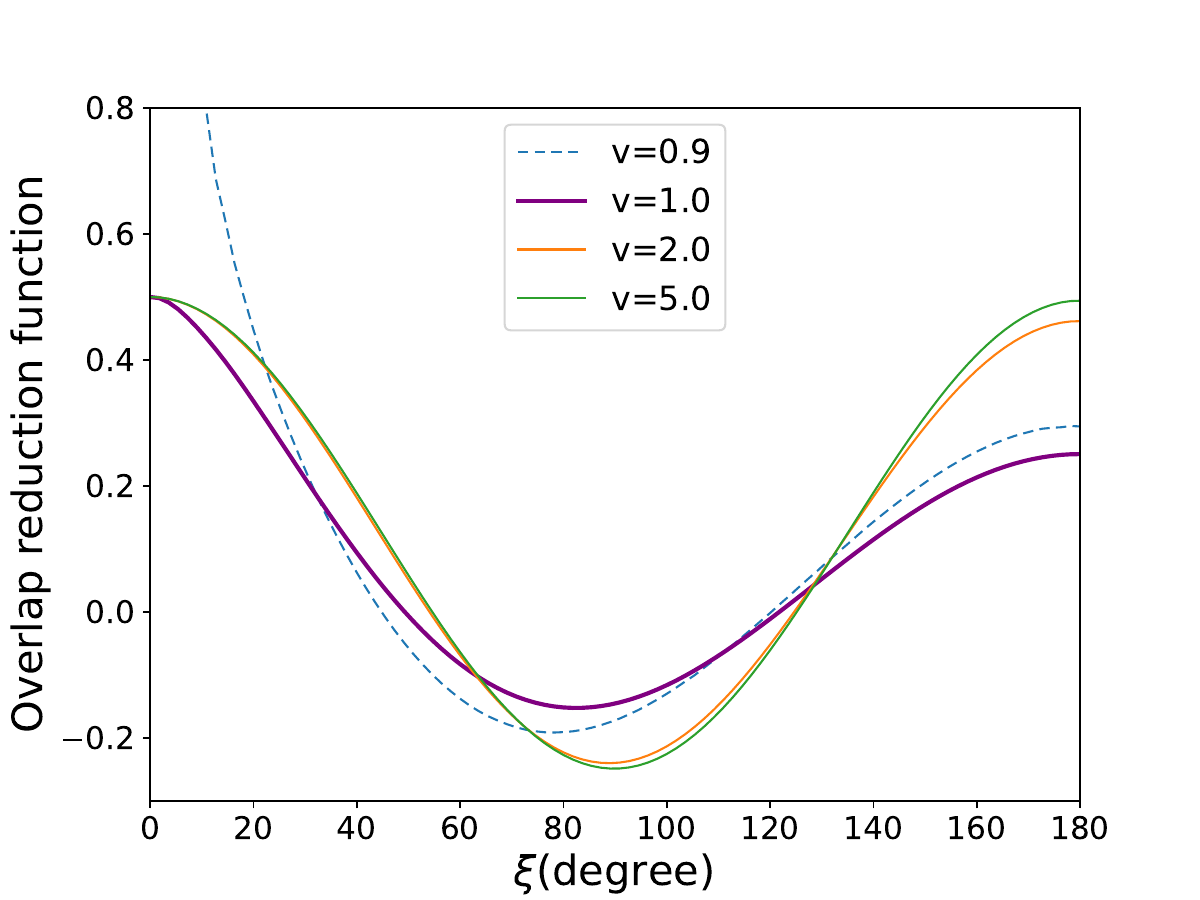}
	\caption{\label{fig:orf} ORF for the SGWB as a function of the angular separation $\xi$ with different GW velocity $v$. Note that we normalize the ORF such that at $\xi = 0$, the value is chosen to be 0.5. For the case with subluminal phase velocity, the ORF tends to diverge at $\xi = 0$. Therefore we choose an arbitrary normalization for comparison.}
\end{figure}

The most general ORF between two pulsars $a $ and $b$ can usually be expressed as
\e
\Gamma_{ab}(f,\xi)=\beta \int d \hat{k} \sum_{A=+,\times} R^{A}_a(f,\hat{k}) R^{A*}_b(f,\hat{k}) ,
\label{orf}
\q
where $\beta$ is the normalization factor. 
The quantity $R^{A}_a(f,\hat{k})$ represents the detector response function for a timing residual measurement. It pertains to a detector with length $L_a$ (namely the distance from the pulsar $a$ to the Earth), sensitive to a plane GW with polarization $A$, propagation direction $\hat{k}$, and frequency $f$.
It can be described as
\e
R^{A}_a(f,\hat{k}) = \frac{1}{i 2\pi f}  \frac{\hat{p}_a^i \hat{p}_a^j e_{ij}^{A}(\hat{k})} {2(1 + \frac{1}{v} \hat{k} \cdot \hat{u})} \( 1 - e^{-i2\pi f L_a (1 + \frac{\hat{k} \cdot \hat{p_a}}{v} )} \) ,
\q
where $\hat{p}_a^i$ is the direction to the pulsar $a$.

A more sophisticated approach to express the ORF is decomposing it into spherical harmonics in the same way that is traditionally applied to the analysis of cosmic microwave background~\cite{Gair:2014rwa,Roebber:2019gha}. In this manner, the ORF is expressed as \citep{Bernardo:2022rif,Liang:2023ary}
\e
\Gamma_{ab}(f,\xi)=\beta \sum_{l=2}^{\infty} (2l+1) \frac{2(l-2)!}{(l+2)!} \vert c_l(f) \vert^2 P_l(\cos \xi_{ab}) ,
\label{orf2}
\q
where $P_l(\cos \xi)$ is the Legendre polynomial and the coefficient $c_l(f)$ is written as \citep{Liang:2023ary}
\begin{widetext}
\e 
c_l(f) = 2i(l+1) \int_{-1}^{1} dx e^{-i\pi fL(1+x/v)} \frac{\sin (\pi fL (1+x/v))}{(1+x/v)} (P_l(x)(-l + (2+l)x^2) - 2xP_{l+1}(x)) ,
\q
\end{widetext}
where $L$ stands for the typical distance of pulsars and the quantity $fL$ is set to $100$ \citep{Anholm:2008wy}. Following \citep{Liang:2023ary}, we can safely ignore the exponential factor when $v \geq 1$ while keeping it in the opposite case.



\section{\label{data}Data and Methodology} 
The NANOGrav 15-year data set includes observations for 68 pulsars, of which 67 pulsars have an observational timespan over 3 years and have been used for the SGWB search \citep{NANOGrav:2023hde}. 
All of these pulsars collectively generate 2211 pairs. To reduce computation cost, we have pre-calculated the ORFs varying with $v$ at these pair separations by interpolating the ORF into a two-dimensional function of velocity $v$ and pair separation $\xi$. Besides the SGWB signal characterised by the ORF obtained above, several other effects also contribute to TOAs, such as the measurement uncertainties of the timing, and the irregularities of the pulsar’s motion and so on \citep{NANOGrav:2023gor}. In practice, these effects should be analysed all together within the timing residuals,
\e
\delta t = M\epsilon + \delta t_{\rm WN} + \delta t_{\rm RN} + \delta t_{\rm SGWB}
\q
where the $M$ is the design matrix, $\epsilon$ is an offset vector of timing model parameters. Here, $\delta t_{\rm WN}$ is the white noise term accounts for the measurement uncertainty of instruments, for which are described by three parameters ``EFAC", ``EQUAD" and ``ECORR" \citep{NANOGrav:2023ctt}. Besides, $\delta t_{\rm RN}$ represents the red noise term from intrinsic noise of pulsar, modeled as a power law with amplitude $A_{\rm RN}$ and index $\gamma_{\rm RN}$ \citep{Cordes:2010fh,NANOGrav:2023ctt}, 
\e
S(f) = \frac{A^2_{\rm RN}}{12 \pi^2}\(\frac{f}{f_{\rm yr}} \)^{-\gamma_{\rm RN}} f^{-3}_{\rm yr} .
\label{S_red}
\q
The correlations between different TOAs, $(t_i , t_j)$, are calculated using the Wiener-Khinchin theorem \citep{NANOGrav:2023ctt}, resulting in the covariance matrix elements
\e
C_{ij}^{\rm RN} = \int df S(f) \cos(2 \pi (t_i - t_j)).
\q
In practice, we employ the ``Fourier-sum" method to model both the red noise and SGWB signal, utilizing Fourier bases $F$ and their associate amplitudes $a$ which are related to the spectral density \Eq{S_red} \citep{Lentati:2012xb}. Following \citep{NANOGrav:2020bcs,NANOGrav:2023gor}, we use frequencies $f_i = i/T$ with the observational timespan $T = 16.03 {\rm yr}$, and set $i=1-30$ for the red noise and $i = 1-14$ for the SGWB signal. To enhance computational efficiency, the stochastic processes are typically assumed to be Gaussian and stationary \citep{Ellis:2014xgh}.
The log likelihood is evaluated as
\e
\ln L(\delta t|\Theta) = -\frac{1}{2} \[ r^{T} C^{-1} r + \ln \det (2 \pi C) \] ,
\q
where $r = \delta t - Fa - M\epsilon$ and $C = \langle r r^{T} \rangle$ is the total covariance matrix. 
Following the Bayesian inference approach adopted by \cite{NANOGrav:2023gor}, the posterior is given as
\e
P(\mathrm{\Theta}|\dt \bbt) \propto L(\dt \bbt|\mathrm{\Theta})\pi(\mathrm{\Theta}),
\q
where $\pi(\mathrm{\Theta})$ is the prior probability distribution. The parameters and their prior distributions needed for the analyses are listed in \Table{tab:params}.

All the aforementioned analyses rely on the JPL Solar System Ephemeris (SSE) DE440 \citep{Park_2021}. We utilize the \texttt{PINT} timing software \citep{pint} to determine the design matrix $M$ for the timing model, employ the \texttt{Enterprise} package \citep{enterprise} to compute the likelihood $L(\delta t|\Theta)$ by marginalizing over the timing model offset parameters $\epsilon$, and utilize the \texttt{PTMCMCSampler} \citep{PTMCMCSampler} package to conduct Markov Chain Monte Carlo (MCMC) sampling for constraining the velocity of the SGWB.

\begin{table*}[htbp]
    \setlength{\tabcolsep}{18pt}
    \caption{List of the parameters and their prior. Here U and log-U represent the uniform and log-uniform distributions, respectively. Here “one parameter for PTA” means the parameter is common in the whole data set, while “one parameter per pulsar” indicates the parameter varies from pulsar to pulsar; the same goes for the case of “one parameter per band/system”.}
    \label{tab:params}
    \begin{tabular}{cccc}
    \hline\hline
    \textbf{parameter}  &  \textbf{description}  &  \textbf{prior}  &  \textbf{comments} \\
    \hline
    \multicolumn{4}{c}{White noise} \\
    $E_k$   &   EFAC per backend/receiver system   &   U$[0, 10]$ & single pulsar analysis only \\
    $Q_k$   &   EQUAD per backend/receiver system  &  log-U$[-8.5, -5]$ & single pulsar analysis only \\
    $J_k$   &   ECORR per backend/receiver system  &  log-U$[-8.5, -5]$ & single pulsar analysis only \\
    \hline
    \multicolumn{4}{c}{Red noise} \\
    $A_{\rm RN}$   &   Red-noise power-law amplitude  &  log-U$[-20, -11]$ & one parameter per pulsar  \\
    $\gamma_{\rm RN}$   &   Red-noise power-law index  &  U$[0, 7]$ & one parameter per pulsar  \\
    \hline
    \multicolumn{4}{c}{Common-spectrum Process} \\
    $v$   &   Velocity of SGWB   &   log-U$[-0.2, 1.0]$ & one parameter per PTA \\
    $A_{\rm GWB}$   &   Power-law amplitude of SGWB  &  log-U$[-18, -11]$ & one parameter per PTA \\
    $\gamma_{\rm GWB}$   &  Power-law index of SGWB  &  U$[0, 7]$ & one parameter per PTA \\
    \hline
    \end{tabular} 
\end{table*}

When conducting the analysis, we initiate noise analyses by solely considering white and red noise for each individual pulsar. Subsequently, we aggregate all 67 pulsars into a whole PTA, fix the white noise parameters to their maximum-likelihood values estimated from the single pulsar noise MCMC chain, and allow red noise parameters to vary simultaneously with the SGWB signal parameters. In signal search among all the pulsars, fixing white noise parameters has negligible impact on the results \citep{EPTA:2015qep}, but can efficiently reduce the computational cost.


\begin{figure}[tbp]
	\centering
	\includegraphics[width=\linewidth]{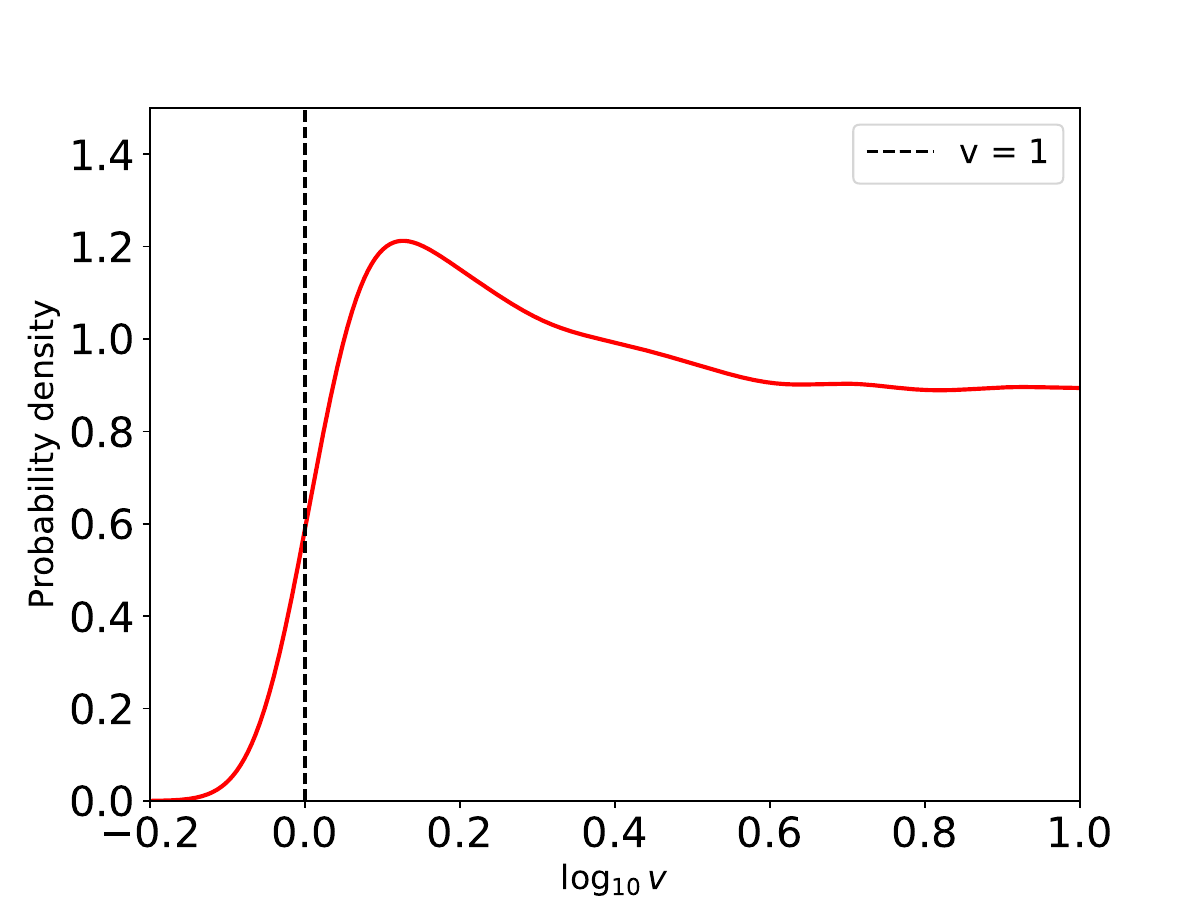}
	\caption{\label{posterior} The posterior distribution for the velocity of GWs.} 
\end{figure}

\section{\label{result}Result and Discussion} 
As previously discussed, the ORF of an SGWB exhibits variations as the velocity of GWs changes. In this work, we derive constraints on the velocity by analyzing these variations. The posterior distribution of the velocity is depicted in \Fig{posterior}, which has been smoothed using the kernel density estimation (KDE) method.
For this analysis, we employ the Gaussian function as the kernel function with a bandwidth set to $0.09$. Additionally, we implement boundary correction \citep{Jones1993, Lewis:2019xzd} for the KDE using the mirroring method.

The posterior of the velocity exhibits the clear lower limit and flattens for velocity larger than the speed of light. As there is not a well-established method for estimating the confidence level (CL) in this particular scenario, we propose a reasonable approach. Specifically, the posterior displays a peak at $\log_{10} v_{\rm peak} \sim 0.127$. Assuming the left side of the peak approximately follows a Gaussian distribution, we use the $1/e^2$ height width to represent the $2\sigma$ CL. This method yields a lower bound of $\log_{10} v \gtrsim -0.059$, or equivalently, $v \gtrsim 0.87$.

The posterior distribution of $v$ is consistent with the variation of ORF with $v$ in \Fig{fig:orf}. Due to significant differences in the ORF with the subluminal case, a natural lower bound can be determined. However, the relatively flat posterior for velocities greater than $1$ indicates that distinguishing the superluminal case from the normal luminal one using the currently detected SGWB remains challenging.
Furthermore, a massive gravity with a non-zero graviton mass seems to correspond to our superluminal velocity case \citep{Bernardo:2023mxc,Bernardo:2023zna}. However, the dispersion relation $\omega = \sqrt{m^2 + \vert k \vert^2 }$ is not equivalent to the dispersion relation we used. Therefore, our approach allows for the exploration of possibilities beyond the commonly assumed massive gravity when introducing variations in the dispersion relation. Its capacity to encompass both the superluminal and subluminal cases also makes our approach unique and generic.






\section*{Acknowledgements}
We acknowledge the use of HPC Cluster of ITP-CAS. QGH is supported by the grants from NSFC (Grant No.~12250010, 11975019, 11991052, 12047503), Key Research Program of Frontier Sciences, CAS, Grant No.~ZDBS-LY-7009, the Key Research Program of the Chinese Academy of Sciences (Grant No.~XDPB15). 
ZCC is supported by the National Natural Science Foundation of China (Grant No.~12247176) and the China Postdoctoral Science Foundation Fellowship No.~2022M710429.

\bibliography{refs}
\end{document}